\documentclass[a4paper,twocolumn,showpacs]{revtex4}

\usepackage{graphicx}
\usepackage{dcolumn}
\usepackage{amsmath}
\usepackage{amssymb}
\usepackage{epsfig}

\usepackage{bm}

\begin{document}
\preprint{APS/123-QED}

\title{Two-dimensional solitons on the surface of magnetic fluids}
\author{Reinhard Richter$^1$, and I.V. Barashenkov$^2$}
\affiliation{$^1$Experimentalphysik V, Universit\"at  Bayreuth,  D--95440 Bayreuth,
Germany}
\affiliation{$^2$University of Cape Town, Rondebosch 7700, South Africa}
\date{\today}

\begin{abstract}
We report an observation of a stable  soliton-like
structure on the surface
of a ferrofluid,  generated by a local perturbation in the hysteretic
regime of the Rosensweig instability.
Unlike other pattern-forming systems with localized 2D structures,
magnetic fluids are characterized by energy conservation; hence
their mechanism of  soliton stabilization is different
from the previously discussed gain/loss balance mechanism.
The radioscopic measurements of the soliton's surface profile suggest
that locking on the underlying periodic structure is instrumental
 in its stabilization.
\end{abstract}

\maketitle

To date, stable solitary waves have been experimentally observed in a 
variety of one-dimensional and quasi-one-dimensional physical systems. In 2D, 
dispersive nonlinear systems are prone to collapse instabilities and hence the 
2D solitons turned out to be more elusive.
(Here we use the term ``soliton" in a broad physical sense, as 
a synonym of localized structure.)
So far, the list of experimentally detectable 2D localized objects was confined mostly
to vortices in superfluids, superconductors, and other media on one hand, and
dissipative solitons in nonequilibrium systems on the other.
While the stability of the former is due to their nontrivial topology, 
the latter come 
into being via the balance of strong dissipation and energy gain. Examples include
current filaments in gas discharge systems \cite{filaments};
oscillons in fluids and granular materials \cite{oscillons};
breathing spots in chemical reactions \cite{chemical}
and feedback and cavity solitons in optics \cite{optics}.
Despite some encouraging theoretical insights,
the question of whether 2D non topological solitons
can arise in {\it conservative\/} systems has remained
open.

In this Letter we report an experimental
observation of a strongly localized, stable stationary
soliton on the surface of magnetic fluid  (MF) in
a stationary magnetic field. MF is a dispersion of
magnetic nanoparticles, and thus
has a high relative permeability $\mu_{\rm r}$\cite{rosensweig1985}.
This is a lossless system;
a horizontal layer of MF in a vertically applied magnetic 
induction ${\boldmath B}$ is characterized by the energy density 
\cite{gailitis1977, friedrichs2001}:
\begin{eqnarray}
{\cal F}[h(x,y)]& = &  \frac{\rho g}{2} h^2(x,y) -\int_0^{h} dz {\boldmath B}\frac{\mu_{\rm r}-1}{2}{\boldmath H_{\rm MF}(x,y,z)}  \nonumber\\
& &+\sigma \sqrt{1+(\partial_x  h(x,y) )^2+(\partial_y   h(x,y) )^2}. 
\label{functional}
\end{eqnarray}
Here $\rho$ and $\sigma$ are the density and surface tension of the MF, $h(x,y)$ the local height
of the liquid layer, and $\boldmath H_{\rm MF}(x,y,z)$ is the magnetic field in the presence of 
the MF. The three terms in Eq.\,(\ref{functional}) represent the hydrostatic, magnetic and
surface energy, respectively.
As the surface profile deviates from the flat reference state,
the first and last term grow whereas the magnetic energy decreases.
For sufficiently large ${\boldmath B}$, this gives 
rise to the normal field, or Rosensweig, instability \cite{cowley1967, rosensweig1985}.

\begin{figure}[htb]
\begin{minipage}{86mm}
\includegraphics[width=75mm]{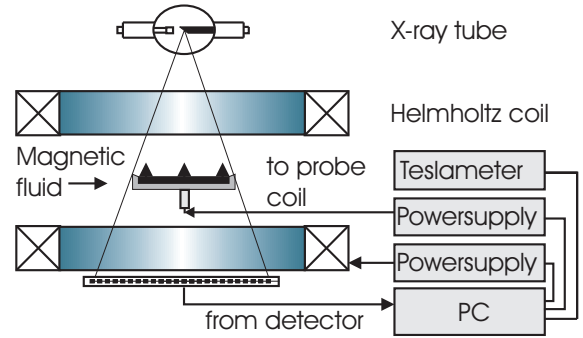}
\end{minipage}
\caption{Sketch of the experimental setup. See text for details. }
\label{setup}
\end{figure}

Our experimental setup is sketched in Fig.\,\ref{setup}.
A Teflon$^{\circledR}$ vessel with the radius  $R=60\,\rm mm$ and depth of 3 mm
\cite{reimann2003} 
is filled with MF up to the brim and placed on the common axis midway between two 
Helmholtz coils.
An x-ray tube is mounted above the center of the vessel at a distance of 1606\,mm.
The radiation transmitted through  the fluid layer and the bottom of the vessel is 
recorded by an x-ray sensitive photodiode array detector (16 bit)
connected to a computer. The full surface relief is then  reconstructed from 
the calibrated radioscopic images. For details see \cite{richter2001}.
The experiments were performed with the magnetic fluid EMG\,901, Lot F121901 AX from 
Ferrotec. Its material parameters have been measured to $\mu_{\rm r}=3.2$, $\rho=1.406 \rm\,g\,cm^{-3}$, 
and $\sigma=25 \pm 0.7 \rm\,mN/m$.

\begin{figure}[htb]
\begin{minipage}{86mm}
\includegraphics[width=75mm]{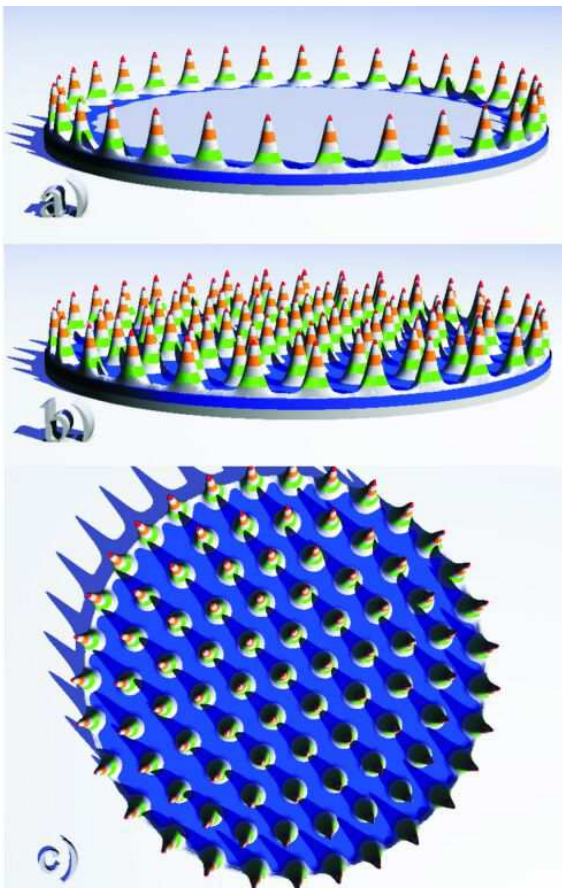}
\end{minipage}
\caption{Surface reliefs as reconstructed from the radioscopic images for a) $B=$ 8.922 mT, 
and b),\,c) 10.407 mT. Each color indicates a layer thickness of 1 mm.}
\label{half_profiles}
\end{figure}

Starting from a flat layer at $B=0 \rm mT$, we increase the external
induction adiabatically in steps of 15 $\rm \mu T$, pausing for 30s after
each increase. 
As shown in Fig.\,\ref{half_profiles}\,(a), a deformation of the surface of the 
liquid appears first at the edge of the vessel. This is due to the magnetic field gradient 
induced by the discontinuous magnetization at the edge of the liquid layer.  
Increasing the induction further gives rise to a fully developed pattern of the 
Rosensweig instability, as shown in (b,c).

\begin{figure}[htb]
\begin{minipage}{86mm} 
\includegraphics[width=86mm]{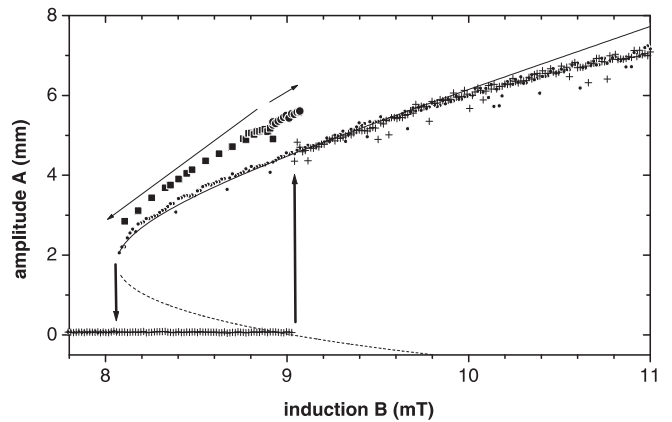}
\end{minipage}
\caption{The amplitude of the pattern for $r<11 \rm mm$ versus the
magnetic induction. The crosses  (dots)  mark the values for increasing (decreasing)
induction respectively. 
The solid (dashed) lines display the least square fit to the roots
\mbox{$A_{\pm}=[\gamma (1+\epsilon) \pm \sqrt{\gamma^2 (1+\epsilon)^2 + 4\,\epsilon g}]/(2g)$},
of the amplitude equation 
$ \epsilon A + \gamma (1+\epsilon) A^2 - g A^3=0$ of Ref. 
\cite{friedrichs2001} with $\gamma=0.281 $ and $g=0.062$. 
The full circles (squares) give the amplitude of the localized spike initiated
at $B=8.91 \rm mT$ for increasing (decreasing) induction respectively.
}
\label{h_vs_B}
\end{figure}

We measured the top-to-bottom height $A$ of the stationary fluid
pattern arising in the adiabatic increase and
decrease of $B$. To avoid the edge-induced imperfections in the character of
the bifurcation, we only consider spikes located within 11 mm from the center 
of the dish. Figure\,\ref{h_vs_B} displays results obtained
for 400 values of $B$. As $B$ is increased, a sudden transition to the
upper branch occurs at $B_c=9.025 \rm mT$. For $B>B_c$, the entire 
surface is covered by a  lattice of liquid spikes, which is hexagonal away from the 
boundary. 
Decreasing $B$, the order parameter $A$ remains on the upper branch
all the way to $B^*=8.076 \rm mT$ where it drops to the flat reference
level. Thus the diagram shows a subcritical bifurcation to hexagons.
The solid and dashed lines display a fit to the roots of the corresponding amplitude equation
of Ref.\,\cite{friedrichs2001}.

\begin{figure}[htb]
\begin{minipage}{86mm}
\includegraphics[width=75mm]{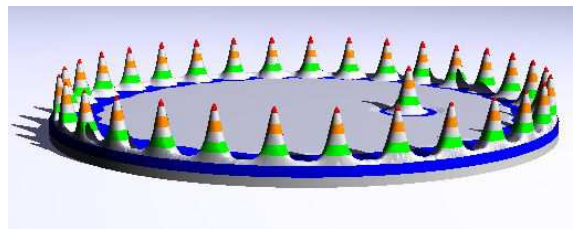}
\end{minipage}

\caption{A single soliton surrounded by the unperturbed magnetic liquid.
The magnetic induction generated by the Helmholtz coils amounts to $B=8.91 \rm mT$.
The amplitude of the local pulse which produced the soliton 
was $B_+= 0.68 \rm mT$ at the bottom of the vessel.
}
\label{LS_1}
\end{figure}

To study the stability of the flat surface to local perturbations (in the hysteretic regime), 
a small air coil with the inner diameter of 8 mm was placed under the center of the vessel 
(see Fig.\,\ref{setup}). This allows to increase, locally, the magnetic induction.  
A local pulse of  $B_+=0.68 \mbox{\rm  mT}$ added to the uniform
field of   $B=8.91 \rm mT$, produces a single stationary spike of fluid, surrounded by a
 circular dip, which does not disperse  after $B_+$ has been turned
 off. Figure\,\ref{LS_1}  presents a measured relief of this radially-symmetric
 state which will be referred to as the soliton. The soliton
 is a stable nondecaying structure; it remained intact for
 days. After its formation at the center of the dish, the soliton
 was often seen to float around (with $v \sim 0.1 \rm mm/s$), until reaching
 an equilibrium position somewhere near the edge of the dish.
 This behavior can be attributed to radial gradients of the magnetic field 
 due to the discontinuous magnetization at the edge and the ring of spikes
 pinned along the perimeter of the dish.

We examined the range of stability of the soliton generated by a pulse with 
$B_+=0.68 \mbox{ \rm  mT}$ added to the uniform induction $B=8.91 \rm mT$. 
Reducing  $B$ adiabatically we measured the corresponding amplitude 
of the soliton (marked by full squares in 
Fig.\,\ref{h_vs_B}). Similarly to the spikes in the hexagonal pattern, the height of the
soliton decreases as $B$ is reduced. The soliton decays for $B<8.09 $mT,
which is close to $B^*= 8.076 $\,mT, the lower 
stability boundary of the hexagonal pattern. As $B$ is increased, the amplitude of
the soliton grows, as indicated in Fig.\,\ref{h_vs_B} by full circles.
At $B=9.055$mT, a sudden transition from the soliton to the fully developed 
Rosensweig pattern occurs.
This value is somewhat larger than $B_c=9.025 \rm \,mT$; 
this is due to the fact that the birth of the soliton produces 
a slight drop in the flat layer thickness, which shifts $B_c$ - similarly to 
Fig.5 in Ref.\cite{friedrichs2001}.

\begin{figure}[tb]
\begin{minipage}{86mm}
\includegraphics[width=65mm]{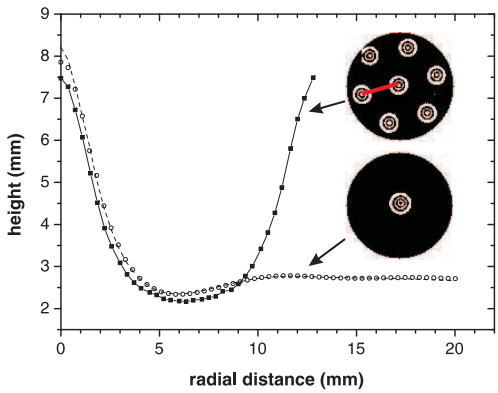}
\end{minipage}
\caption{
The filled squares mark the profile of one period of the hexagonal pattern,
measured at $B= 9.07\rm mT$ in the center of the vessel; $r<8.8$mm.
Azimuthally averaged height profiles of two different solitons, measured at
the same induction  are depicted by open symbols (one) and a dashed line (the other).
}
\label{height_profile}
\end{figure}

In order to illustrate the robustness of the soliton's shape we show in
Fig.\,\ref{height_profile} the azimuthally averaged profiles  of
two different solitons, produced in two separate experiments
at $B=9.07 \mbox{\rm mT }$. The profiles are
practically indistinguishable.
Also plotted are two half-periods of the
corresponding hexagonal lattice.
In agreement with Fig.\,\ref{h_vs_B},  the soliton is about
$1 \mbox{mm}$  taller than the spikes of the lattice.
This may be attributed to the fact that the spikes emerge  simultaneously, and thus 
have to share the liquid available.
However, the width of the soliton is exactly equal 
to the period of the lattice. Therefore, there is 
a preferred wavelength in the system, defined by the lattice,
to which the soliton locks. As we show now, 
this width locking is the central part of the 
soliton's stabilization mechanism.

Consider  the dispersion relation
of a semi-infinite layer of  inviscid MF \cite{cowley1967},
\begin{equation}
 \omega^2 = gk - \mu_0 \mu_r
 \frac{(\mu_r - 1)^2}{\mu_r +1} \frac{1}{\rho}H^2k^2 + \frac{
 \sigma}{\rho} k^3 .
  \label{dispersion}
  \end{equation}
Here $\omega$ is the frequency, $k=|\vec k|$ the wavenumber,
$H$ the strength  of the external magnetic field, and $\mu_0$ the magnetic
field constant.
While the first and the third terms account for the
gravity and capillary effects and are common for all fluids,
the second term is specific just for the MF. As $H$ is
increased above $H_c$, where
$ H_c^2=\frac{2\,(\mu_r+1)}{\mu(\mu_r-1)^2 }\ (\rho g \sigma)^{1/2}$,
 a band
of  wavenumbers with $\omega^2<0$ appears around
   $k_c=(\rho g / \sigma)^{1/2}$ and the flat state
   loses its stability to  the hexagonal
   pattern. Since the soliton should decay 
   to the flat surface as ${\vec x}^2 \to \infty$, 
   there can be no stable solitons for $H>H_c$.
   
Let now  $H<H_c$. On the qualitative level,  
our system can be modelled by
 a conservative analogue of the
 Swift-Hohenberg equation:
\begin{equation}
{\ddot u} +(k_0^2+\nabla^2)^2 u+au =3bu^2 -2cu^3; \quad b,c>0.
\label{SHE}
\end{equation}
Eq.(\ref{SHE}) can be used as a model since (a) it
 has a nonmonotonic dispersion relation
$\omega^2=a+k_0^4-2k_0^2k^2+k^4$, where, as in 
(2), the destabilizing $k^2$-term is opposed by a higher
power of $k$; and (b) it has a symmetry-breaking hysteretic nonlinearity 
which was shown to provide a fairly accurate 
approximation of the amplitude of the hexagonal pattern
\cite{friedrichs2001} (see Fig.\ref{h_vs_B}.) 
We have verified, numerically, that Eq.(\ref{SHE}) does indeed have
a stable stationary radially-symmetric soliton solution
coexisting with stable hexagons
in a broad parameter range. Its stability can be
explained by a Derrick-type argument for the corresponding energy 
functional, 
\begin{equation}
E=  \int [
{\dot u}^2 +V(u)-2k_0^2 (\nabla u)^2+(\nabla^2 u)^2
                      ] d^2x.
\label{energy}
\end{equation}
Here $V(u)=(a+k_0^4)u^2-2bu^3+cu^4$. A scaling perturbation
 $u({\vec x}) \to u(\kappa {\vec x})$ takes the stationary energy to
 \[
 E(\kappa^2)= \frac{1}{\kappa^2} \int V d^2x -
 2k_0^2 \int (\nabla u)^2 d^2x
 + \kappa^2 \int (\nabla^2 u)^2 d^2x.
 \]
The first term
(nonlinearity) opposes the dispersive broadening of the soliton
(for which $\kappa \to 0$)
while the last one prevents the nonlinear blow-up
(for which $\kappa \to \infty$). In a similar way,
 the first and last term in (\ref{dispersion}) make contributions to
the energy which scale as $\kappa^{-1}$ and $\kappa$, respectively.
The first term (along with the nonlinearity) 
opposes  the spreading and
 the last one arrests the  blow-up. 
 Next, setting the derivative $(dE/d\kappa^2)_{\kappa=1}$ to zero,
 gives
 $\int V d^2x=\int (\nabla^2 u)^2 d^2x$. Using this
 relation, the second derivative, $ d^2 E/d (\kappa^2)^2$, is calculated to be 
 $2\int (\nabla^2 u)^2 d^2x>0$, which means that the soliton renders the energy 
 minimum. If Eq.(\ref{SHE}) did not
include the higher-derivative term, the energy would not 
have a nontrivial minimum.  The introduction of a higher derivative 
(or, equivalently, the nonmonotonicity of the dispersion curve)
sets a preferential wavelength in the system --- to which the 
soliton locks and stabilizes.  
 A similar stabilization mechanism was discussed before
in the context of the wave front locking
\cite{pomeau1986}; see also \cite{sakaguchi1997}.

\begin{figure}[bht]
\begin{minipage}{86mm}
\includegraphics[width=65mm]{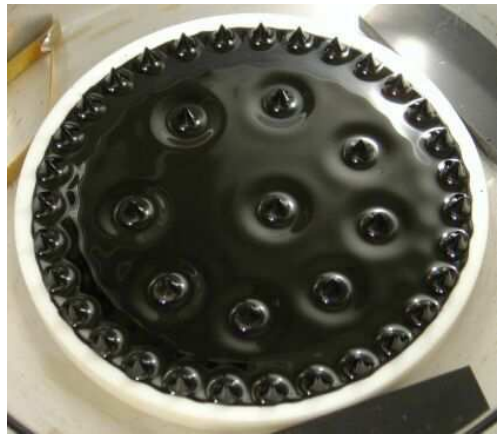}
\end{minipage}
\caption{Nine solitons at  $B=8.91 \rm mT$.}
\label{LS_9}
\end{figure}

\begin{figure}[htb]
\begin{minipage}{86mm}
\includegraphics[width=65mm]{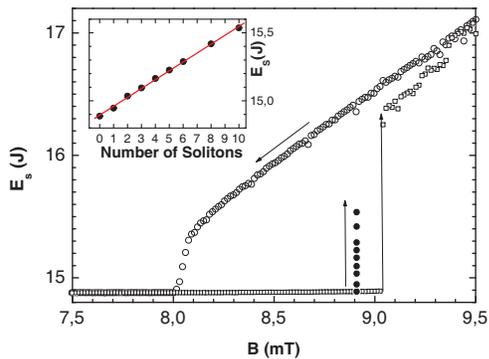}
\end{minipage}
\caption{The surface energy of the liquid layer for increasing (open squares) 
and decreasing (circles) magnetic induction.
The full circles mark the increase of $E_{\rm s}$ through the successive 
generation of solitons at  $B=8.91 \mbox{\rm mT}$ (see also inset). 
To reduce the influence of the perimeter spikes, only the area $r<0.88\,R$ was  
covered, where $R$ is the radius of the vessel.
}
\label{E_surface}
\end{figure}

Applying, repeatedly, pulses of $B_+$ and allowing the newly 
born solitons to drift away from the site of the probe coil, 
we were able to generate two, three, and more solitons. 
Figure \ref{LS_9} presents an example of a 9-soliton 
configuration, with only one remaining at the center. 
In this way, it is possible to increase the surface energy
of the liquid layer in steps. 
This is illustrated in Fig.\,\ref{E_surface} which also shows 
the surface energy of the Rosensweig pattern as a hysteretic 
function of $B$.
Thus, one can reach the region between the two branches of 
this function which is not accessible for the standard 
Rosensweig instability.

\begin{figure}[h]
\begin{minipage}{86mm}
\includegraphics[width=86mm]{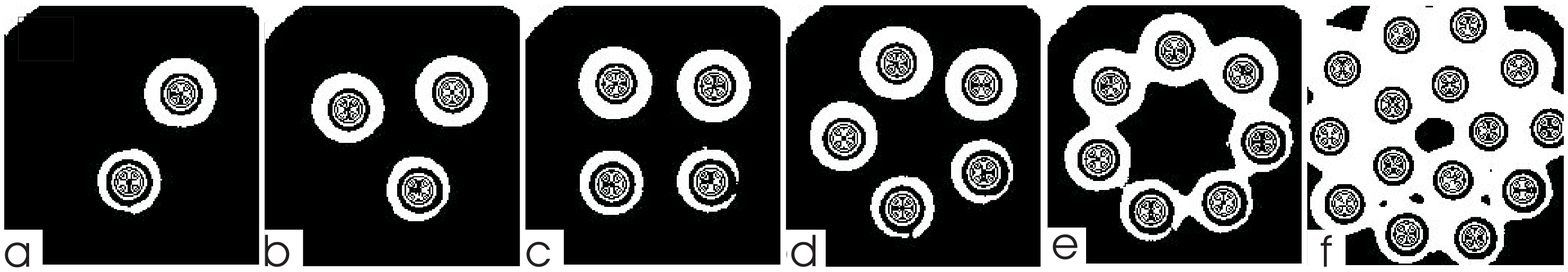}
\end{minipage}
\caption{The solitons can form molecule-like clusters:
(a) di-, (b) tri-, (c) tetra-, (d) penta-, (e) hepta-, and (f) oligomers. The height 
is indicated by switching between black and white after each mm. Here $B= 8.71\rm mT$;
each panel covers  the area of $(87\rm mm)^2$.
}
\label{molecules}
\end{figure}

Could solitons serve as building blocks in the formation of
periodic patterns? We have observed that, if additional
care is taken to suppress the edge-induced inhomogeneity of
the magnetization,
solitons can form molecule-like clusters
(Fig.\,\ref{molecules}.) This may seem to contradict the
repulsive nature of the dipole-dipole interaction; however,
there is a simple mechanism that can account for the binding.
Indeed, each soliton is surrounded by concentric
dips representing ring-like regions of depleted magnetic induction.
The innermost, deepest, ring is clearly visible in 
Figs.\,\ref{LS_1},\ref{height_profile}, and \ref{LS_9};
a higher-resolution measurement allows to discern another,
shallower dip of larger radius. The dips create a  potential relief
 which may capture the partner
soliton(s). As the density of solitons grows, the
multisoliton cluster evolves towards the hexagonal Rosensweig
lattice. It still remains to be understood
whether the cluster-lattice transition requires additional excitation
energy.

In conclusion, we found stable 2D solitons on the
surface of a ferrofluid in the hysteretic regime of the
Rosensweig instability.
These objects are easy to generate and control
and they are easily set in motion; this
opens ways for studying their binding and scattering.
Due to the {\em conservative nature}
of the ferrosolitons, and unlike the localized structures
 observed previously in
dissipative systems,  the balance of dissipation and energy gain
plays no role in their stabilization.
Instead, we suggest a stabilization mechanism which appeals to the locking
of the soliton  to the  wavelength imposed by the
nonmonotonic dispersion relation.
This mechanism can also be at work in other conservative
systems with preferred wavelengths, e.g. in electrostatics and
elasticity \cite{taylor1965}.

   We thank I.\,Rehberg, W.\,Pesch and R.\,Friedrich for 
discussions and K.\,Staliunas for providing a  code for Eq.(\ref{SHE}).
Support by NRF of South Africa, grant 205723, and 
Deutsche For\-schungs\-gemein\-schaft,  grants HBFG\,051-201 and Ri 1054/1-4
is gratefully acknowledged.

\end{document}